\documentstyle[12pt,epsfig]{article}
\evensidemargin=\oddsidemargin
\renewcommand{\baselinestretch}{0.93}

\renewenvironment{thebibliography}[1]
    {\begin{list}{\arabic{enumi}.}
    {\usecounter{enumi}\setlength{\parsep}{0pt}
\setlength{\leftmargin 1.25cm}{\rightmargin 0pt}
     \setlength{\itemsep}{0pt} \settowidth
    {\labelwidth}{#1.}\sloppy}}{\end{list}}
           
\begin{document}
\newcommand{\lae}{\stackrel{<}{\sim}}
\newcommand{\gae}{\stackrel{>}{\sim}}
%
\def\be{\begin{equation}}
\def\ee{\end{equation}}
\def\bea{\begin{eqnarray}}
\def\eea{\end{eqnarray}}
\def\CPbar{\hbox{{\rm CP}\hskip-1.80em{/}}}
\def\D0{D\O~}
\def\pbarp{ \bar{{\rm p}} {\rm p} }
\def\pp{ {\rm p} {\rm p} }
\def\ifb{ {\rm fb}^{-1} }
\def\del{\partial }
\def\ra{\rightarrow}
%
\def\today{\number\day
           \space\ifcase\month\or
             January\or February\or March\or April\or May\or June\or
             July\or August\or September\or October\or November\or  
December\fi           \space\number\year}

\def\dis{\displaystyle}
\def\be{\begin{equation}}
\def\ee{\end{equation}}
\def\ra{\rightarrow}
\def\hhs{\hat{s}}
\def\hht{\hat{t}}
\def\hhu{\hat{u}}
\def\c{c_{\theta}}
\def\ax{\alpha}
\def\a{\alpha_4}
\def\ac{\alpha_{46}}
\def\ab{\alpha_{45}}
\def\b{\alpha_5}
\def\bd{\alpha_{57}}
\def\cde{\alpha_{6710}}
\def\cc{\alpha_6}
\def\d{\alpha_7}
\def\cd{\alpha_{67}}
\def\e{\alpha_{10}}
\def\hs{\hat{s}}
\def\hhs{\hat{s}}
\def\hht{\hat{t}}
\def\hu{\hat{u}}
\def\cut{\displaystyle\left(\frac{\Lambda}{\Lambda_0}\right)^2}
\def\f{\frac}
\def\ETslash{\not{\hbox{\kern-4pt $E_T$}}}
\def\mynot#1{\not{}{\mskip-3.5mu}#1 }
\def\sss{\scriptscriptstyle }
\def\ra{\rightarrow}
\def\to{\rightarrow}
\def\To{\Rightarrow}
\def\d{{\rm d}}
\def\M {{\cal M}}
\def\WW {W_L W_L}
\def\MWW {M_{WW} }
\def\WWWW {W_L W_L \ra W_L W_L}
\def\PPPP {\phi \phi \ra \phi \phi}
\def\WPWM{ W^+(\ra \ell^+\nu) W^-(\ra q_1 \bar q_2) }
\def\WPZ{ W^+(\ra \ell^+ \nu) Z^0(\ra q \bar q) }
\def\WPWP{ W^+(\ra \ell^+ \nu) W^+(\ra \ell^+ \nu) }
\def\etc{ {\it etc.}}
\def\dis{\displaystyle}
\def\f{\frac}
\def\OO{\cal O}
\def\nor{\normalsize}

\thispagestyle{empty}


\rightline{UCD-97-22}
\rightline{MADPH-97-1025}
\rightline{MSUHEP-70930}
\rightline{hep-ph/9711429}
\rightline{November 1997}
\vspace*{0.5cm}

\setcounter{footnote}{0}
\renewcommand{\thefootnote}{\fnsymbol{footnote}}

\begin{center}
 {\Large {\bf Quartic Gauge Boson Couplings at Linear  
Colliders}}\\[0.14cm]
{\normalsize
--- {\bf Interplay of $WWZ/ZZZ$ Production and $WW$-Fusion} --- }

\vspace{1cm}
 {\sc Tao~Han}~$^{1,2}$ \footnote{Email:~THan@Ucdhep.Ucdavis.Edu},~~~~~
 {\sc Hong-Jian~He~}$^{3,4}$ \footnote{
Email:~HJHe@Desy.De~~HJHe@Pa.Msu.Edu},~~~~~
 {\sc C.--P.~Yuan}~$^4$ \footnote{Email:~Yuan@Pa.Msu.Edu}\\[0.25cm]

       {\it $^1$Department of Physics, University of California}\\
       {\it Davis, California 95616, USA }\\
       {\it $^2$Department of Physics, University of Wisconsin}\\
       {\it Madison, Wisconsin 53706, USA }\\[0.3cm]
       {\it $^3$Theory Division, Deutsches Elektronen-Synchrotron DESY}\\
 {\it D-22603 Hamburg, Germany}\\[0.3cm] 
          {\it $^4$Department of Physics and Astronomy,
 Michigan State University}\\
 {\it East Lansing, Michigan 48824, USA}

\vspace{1.9cm}
\begin{abstract}
\noindent
We study new physics effects to the quartic gauge boson couplings formulated
by the electroweak chiral Lagrangian. Five next-to-leading order 
operators characterize the anomalous quartic gauge interactions
which involve pure Goldstone boson dynamics for the electroweak
symmetry breaking.  We estimate the typical size of these couplings
in different strongly-interacting models and examine the sensitivity
to directly probing them via the $WWZ/ZZZ$ triple gauge boson production 
at the high energy linear colliders. 
The important roles of polarized $e^-$ and $e^+$ beams are stressed. 
We then compare the results with those from the $W$-pair production of 
the $WW$-fusion processes, 
and analyze the interplay of these two production mechanisms for
an improved probe of the quartic gauge boson interactions. \\[0.4cm]
\noindent PACS number(s): 11.30.Qc, 11.15.Ex, 12.15.Ji, 14.70.--e \\[0.2cm]
\noindent Physics Letters B (1998), in press.

%
\end{abstract}
\end{center}

\baselineskip20pt   

\newpage
\renewcommand{\baselinestretch}{0.95}
\setcounter{footnote}{0}
\renewcommand{\thefootnote}{\arabic{footnote}}
\setcounter{page}{1}

\noindent
{\normalsize \bf 1.~Electroweak Symmetry Breaking
         and Quartic Gauge Boson Couplings}
\vspace*{0.3cm}

The validity of the Standard Model (SM) has been
tested to a great precision up to
the scale of $O(100)$~GeV \cite{search}, while
the electroweak symmetry breaking (EWSB)
mechanism remains undetermined. If the EWSB sector is
weakly coupled, such as in the supersymmetric theories,
the effects of new physics at higher mass scales generally
decouple from the low energy phenomena \cite{decouple}.
If the EWSB sector is strongly interacting instead, in lack
of phenomenologically viable models for the strong
dynamics \cite{kenlane}, the effective field theory
approach \cite{georgi} provides the most general and economic
description.
In this case, the non-decoupling property of the electroweak
sector enforces the new physics scale ($\Lambda$) to lie below 
or at $4\pi v\simeq 3.1$~TeV, where $v$ is
the vacuum expectation value characterizing the EWSB.

Below the possible heavy resonances \cite{B-Y} at the scale $\Lambda$, 
the new physics effects in the non-decoupling scenario
can be systematically parametrized
as the next-to-leading order (NLO) operators of the electroweak
chiral Lagrangian (EWCL) \cite{EWCL}.
There are in total fifteen NLO bosonic operators,
which can modify the electroweak gauge boson interactions
and induce the so-called ``anomalous couplings'' \cite{anomal}.
The contributions of all these
NLO operators to various high energy processes have
been recently classified by means of a global power counting
analysis \cite{global}. Five operators contain quartic gauge
couplings (QGCs) which can involve the pure Goldstone boson interactions
(according to the equivalence theorem~\cite{et}) and are thus of special
importance for probing the EWSB. 
They are summarized as follows \cite{EWCL}:\\[-0.5cm]
\be
\begin{array}{lc}
\left\{
\begin{array}{lll}
{\cal L}_4 & \, = & \ell_4  \left(\frac{v}{\Lambda}\right)^2
              [{\rm Tr}({\cal V}_{\mu}{\cal V}_\nu )]^2 ~,\\ [0.2cm] 
{\cal L}_5 & \, = & \ell_5 \left(\frac{v}{\Lambda}\right)^2
              [{\rm Tr}({\cal V}_{\mu}{\cal V}^\mu )]^2 ~;
\end{array}\right\}~~
& (SU(2)_c{\rm -conserving})  \\[0.65cm]
\left\{
\begin{array}{lll}
{\cal L}_6 & = &\ell_6\left(\frac{v}{\Lambda}\right)^2
[{\rm Tr}({\cal V}_{\mu}{\cal V}_\nu )]
{\rm Tr}({\cal T}{\cal V}^\mu){\rm Tr}({\cal T}{\cal V}^\nu) ~,\\[0.2cm]
{\cal L}_7 & = & \ell_7\left(\frac{v}{\Lambda}\right)^2
[{\rm Tr}({\cal V}_\mu{\cal V}^\mu )]
{\rm Tr}({\cal T}{\cal V}_\nu){\rm Tr}({\cal T}{\cal V}^\nu) ~,\\[0.2cm]
{\cal L}_{10} & = &\ell_{10}\left(\frac{v}{\Lambda}\right)^2
\frac{1}{2}
[{\rm Tr}({\cal T}{\cal V}^\mu){\rm Tr}({\cal T}{\cal V}^{\nu})]^2 ~;
\end{array}\right\}~~
&  (SU(2)_c{\rm -violating})
\end{array}
\label{eq:Leff}
\ee
where {\small $~{\cal V}_\mu \equiv (D_\mu U)U^\dagger $,
$D_{\mu}U  = \partial_{\mu}U + {\bf W}_{\mu}U -U{\bf B}_{\mu}$,
${\bf W}_{\mu}\equiv ig W^a_{\mu}\tau^a/{2}$,
${\bf B}_{\mu}\equiv ig^{\prime}B_{\mu}\tau^3/{2}$,
$~U  =  \exp [i\tau^a\pi^a/v ]~$}, and
{\small $~{\cal T}\equiv U\tau_3 U^\dagger ~$};
$\pi^a$ is the would-be Goldstone boson field
and $W^a_\mu, B_\mu$ the gauge fields of the $SU(2)\otimes U(1)$ group.
Note that the operators ${\cal L}_{6,7,10}$ violate the custodial 
$SU(2)_c$ symmetry even in the limit $g' \rightarrow 0$.
No current experiment ever reaches the sufficient energy threshold
to directly probe these five operators which involve only QGCs.
Some roughly estimated bounds were derived through the  
1-loop contributions\footnote{
This is already at the level of $\f{1}{16\pi^2}\f{v^2}{\Lambda^2}\lae 
\f{v^4}{\Lambda^4}$, i.e., of the two-loop order.}
to the $Z$-pole measurements by keeping only the logarithmic terms. 
For $\Lambda =2$~TeV and keeping only one parameter to be non-zero at a time, 
the $90\%$~C.L. bounds are  \cite{LEP-QGC}:
\be
\begin{array}{c}
-4 \leq \ell_4 \leq 20~,~~~~-10 \leq \ell_5 \leq 50 ~;\\
-0.7 \leq \ell_6 \leq 4~, ~~~~-5 \leq \ell_7 \leq 26 ~,~~~~
-0.7 \leq \ell_{10} \leq 3 ~.
\end{array}
\label{eq:LEPbound}
\ee
%
Note that in (\ref{eq:Leff}) the dependence on $v$ and $\Lambda$
is factorized out so that the dimensionless
coefficient $\ell_n$  (for the operator ${\cal L}_n$) is
naturally of $O(1)$ \cite{georgi}. The bounds in (\ref{eq:LEPbound})
on the $SU(2)_c$ symmetric parameters
$\ell_{4,5}$ are about an order of magnitude above their
natural size; while the allowed range for the $SU(2)_c$-breaking 
parameters $\ell_{6-10}$ is about a factor of $O(10-100)$
larger than that for $\ell_0=\f{\Lambda^2}{2v^2}\Delta\rho
\left(=\f{\Lambda^2}{2v^2}\alpha T\right)$ derived from
the $\rho$ (or $T$~\cite{stu}) parameter:
$~0.052 \leq \ell_0 \leq 0.12$~\cite{global},
for the same $\Lambda$ and confidence level.
To directly test the EWSB dynamics, it is therefore
important to probe these QGCs at future 
high energy scattering processes where their contributions can be greatly 
enhanced due to the sensitive power-dependence on 
the scattering energy \cite{global}.

In this Letter, we study the sensitivity of the high energy $e^+e^-$ linear
colliders (LCs) with polarized beams to testing these QGCs via
the $WWZ/ZZZ$-production processes. The interplay of these processes
with the $W$-pair production from $WW$-fusion is analyzed.

\vspace{0.4cm}
\noindent
{\normalsize\bf  2.~Relations to Strongly-Interacting Models }
\vspace*{0.25cm}

Although the fundamental theory behind the effective EWCL
is yet unknown, it is important to examine how a typical
strongly-interacting electroweak model
would contribute to these EWSB parameters. We shall
concentrate on the quartic gauge couplings
($\ell_n$'s) in (\ref{eq:Leff})
and consider a few representative models which may contain an isosinglet
scalar $S$, an isotriplet vector $V^a_\mu$,
an isotriplet axial vector $A^a_\mu$,
and new heavy chiral fermions.

\noindent
\underline{\bf A Non-SM Singlet Scalar}

Up to dimension-$4$
and including both the $SU(2)_c$ conserving and breaking effects, we
can write down the most general Lagrangian for an isosinglet scalar: 
\be
{\small
\begin{array}{ll}
{\cal L}_{\rm eff}^{S} ~= &
\dis\f{1}{2}\left[\partial^\mu S\partial_\mu S -M_S^2S^2\right] - V(S)\\
&  -\dis\left[\f{\kappa_s}{2}vS+\f{\kappa_s^{\prime}}{4} S^2\right]
  {\rm Tr}\left[{\cal V}_\mu {\cal V}^\mu \right]
  -\left[\f{\tilde{\kappa}_s}{2}vS+\f{\tilde{\kappa}_s^{\prime}}{4}  
S^2\right]
   \left[{\rm Tr} {\cal T}{\cal V}_\mu \right]^2
\end{array}
\label{eq:scalar}
}
\ee
where $V(S)$ contains only the self-interactions of the scalar $S$. 
The SM Higgs scalar corresponds to a special parameter-choice:
$~\kappa_s=\kappa_s'=1,~ \tilde{\kappa}_s=\tilde{\kappa}_s'=0~$ and 
{\small $~V(S)=V(S)_{\rm SM}~$.}
When the scalar mass ($M_S$) is heavy,  $S$ can be integrated out
from the low energy spectrum with its effects formulated in the heavy
mass expansion:
\be
{\small
\widehat{\cal L}_{\rm eff}^S ~=~ \dis\f{v^2}{8M_S^2}
\left[ \kappa_s {\rm Tr}\left({\cal V}_\mu{\cal V}^\mu\right) +
       \tilde{\kappa}_s {\rm Tr}\left({\cal T}{\cal V}_\mu\right)^2  
\right]^2
        + O\left(\f{1}{M_S^4}\right)
\label{eq:S-Leff}
}
\ee
After identifying {\small $\Lambda =M_S$~} and comparing with 
(\ref{eq:Leff}), we obtain
\be
{\small
\ell_4^s =0~,~~~\ell_5^s =\f{\kappa_s^2}{8} \geq 0~;~~~~~
\ell_6^s =0~,~~~\ell_7^s=\f{\kappa_s\tilde{\kappa}_s}{4}~,~~
\ell_{10}^s =\f{\tilde{\kappa}_s^2}{8} \geq 0 ~.
\label{eq:S-pattern}
}
\ee
In the SM, the only non-zero coupling at the tree level is
$~(\ell_5^s)_{\rm SM}=0.125$;~ while for a different coupling 
choice: $\kappa_s=-\tilde{\kappa}_s=2$, we would have $~\ell_5^s=0.5,~
\ell^s_7=-2\ell^s_{10}=-1$.

\noindent
\underline{\bf Vector and Axial-Vector Bosons}

The LEP-I $Z$-pole measurement on the $S$-parameter
disfavors the naive QCD-like technicolor dynamics~\cite{stu},
where the vector particle $\rho_{\rm TC}$ (technirho) 
is the lowest new resonance in the TeV regime. 
In modeling the non-QCD-like dynamics,
it was suggested \cite{bess} that the coexistence of
nearly degenerate vector and axial-vector bosons may provide
sufficient cancellation for avoiding large corrections to the $S$-parameter.
We thus consider both the vector fields $V^a_\mu$ 
and the axial-vector fields $A^a_\mu$,
as isospin triplets of the custodial $SU(2)_c$. 
Under the global $SU(2)_c$, $V$ and $A$ transform as\\[-0.5cm]
\be
{\small
\widehat{V}_\mu \To \widehat{V}_\mu^{\prime}
                   =\Sigma_v\widehat{V}_\mu\Sigma_v^\dagger ~,~~~~~
\widehat{A}_\mu \To \widehat{A}_\mu^{\prime}
                   =\Sigma_v\widehat{A}_\mu\Sigma_v^\dagger ~,
\label{eq:VA-transf}
}
\ee
where
{\small $\widehat{V}_\mu \equiv V_\mu^a\tau^a/2,~
\widehat{A}_\mu \equiv A_\mu^a\tau^a/2$, }
and {\small $\Sigma_v\in SU(2)_c~$.}
If $V$ and $A$ are further regarded as gauge fields of a new
local hidden symmetry group
{\small ${\cal H}=SU(2)_L'\otimes SU(2)_R'$}
(with a discrete left-right parity) \cite{bess},
we can write down the following general
Lagrangian (up to two derivatives), in the
{\it unitary gauge}\footnote{%
By ``unitary gauge'' we mean a gauge containing no new Goldstone
bosons other than the three for generating the longitudinal
components of the $W,Z$ bosons.
In fact, it is not necessary to introduce such a new
local symmetry ${\cal H}$ for $V,A$~\cite{georgi1}
since ${\cal H}$ has to be broken anyway and
$V,A$  can be treated as matter fields~\cite{CCWZ}. The hidden
local symmetry requires the coefficients of the terms $-J^V_\mu V^\mu$
and $V_\mu V^\mu$ to be the same, due to the additional assumption
about that new local group ${\cal H}$.} of the group ${\cal H}$
and with both $SU(2)_c$-conserving and -breaking effects included, 
\be
{\small
\begin{array}{l}
{\cal L}_{\rm eff}^{VA} =
{\cal L}^{VA}_{\rm kinetic} -\dis \f{v^2}{4}\left[
\kappa_0 {\rm Tr}{\cal V}_\mu^2 +
\kappa_1 {\rm Tr}\left(J^V_\mu -2V_\mu\right)^2+
\kappa_2 {\rm Tr}\left(J^A_\mu+2A_\mu\right)^2+
\kappa_3 {\rm Tr}A_\mu^2\right]  \\[0.25cm]
~~~~~~\left. \hspace{1.5cm}
+\tilde{\kappa}_1\left[{\rm Tr}\tau^3 (J^V_\mu -2V_\mu )\right]^2
+\tilde{\kappa}_2\left[{\rm Tr}\tau^3 (J^A_\mu +2A_\mu )\right]^2
+\tilde{\kappa}_3\left[{\rm Tr}\tau^3 A\right]^2 \right]
\end{array}
\label{eq:VA}
}
\ee
with
\vspace{-0.4cm}
\be
{\small
\begin{array}{ll}
\left\{\begin{array}{l} J^V_\mu =J^L_\mu +J^R_\mu \\[0.2cm]
                        J^A_\mu =J^L_\mu -J^R_\mu
       \end{array} \right.
&~~~~~
\left\{\begin{array}{l}
J^L_\mu =\xi^\dagger D^L_\mu\xi
        =\xi^\dagger \left(\partial_\mu + {\bf W}_\mu\right)\xi\\[0.2cm]
J^R_\mu =\xi D^R_\mu\xi^\dagger
        =\xi\left(\partial_\mu +{\bf B}_\mu\right)\xi^\dagger 
\end{array}  \right.
\end{array}
\label{eq:VA-current}
}
\ee
and {\small  $~V_\mu\equiv i\tilde{g}\widehat{V}_\mu  
=i\tilde{g}V^a_\mu\tau^a/2,
         ~A_\mu\equiv i\tilde{g}\widehat{A}_\mu
=i\tilde{g}A^a_\mu\tau^a/2$,}
and {\small  $\xi \equiv U^{\f{1}{2}}$,}
where $\tilde g$ is the gauge coupling of the group ${\cal H}$.
Among the above new parameters ($\kappa_n$'s), $\kappa_0$ is  
determined by normalizing the Goldstone boson kinetic term:
$~\kappa_0 = -4\kappa_2\kappa_3/(4\kappa_2+\kappa_3)~$.
After eliminating the $V$ and $A$ fields in the heavy mass expansion,
we obtain   
\be 
{\small
\begin{array}{ll}
\widehat{\cal L}_{\rm eff}^{VA} & 
=~ \dis\f{(\eta^2-1)^2+16\tilde{\eta}^2}{8\tilde{g}^2}
\left[
\left({\rm Tr}{\cal V}_\mu{\cal V}_\nu\right)^2  -
\left({\rm Tr}{\cal V}_\mu{\cal V}^\mu\right)^2 \right]+ \\[0.3cm]
& \dis\f{\tilde{\eta}\left(4(3-\eta^2)\tilde{\eta}
+ (1-\eta^2)\eta\right)}
{8\tilde{g}^2}\left[ 
 {\rm Tr}{\cal V}_\mu^2\left({\rm Tr}{\cal T}{\cal V}_\nu\right)^2
-{\rm Tr}\left({\cal V}_\mu{\cal V}_\nu\right)
 {\rm Tr}\left({\cal T}{\cal V}^\mu\right)
 {\rm Tr}\left({\cal T}{\cal V}^\nu\right)
\right] 
+ O\left(\f{1}{M_{V,A}^4}\right) \nonumber
\end{array}
\label{eq:VA-Leff}
}
\vspace*{-0.5cm}
\ee
which contributes to $\ell_{n}$ as follows:
\be
{\small
\begin{array}{ll}
\left\{\begin{array}{l}  \ell_4=\ell_4^v+\ell_4^a \\[0.3cm]
                    \ell_5=\ell_5^v+\ell_5^a \\[0.3cm]
                    \ell_6=\ell_6^v+\ell_6^a \\[0.3cm]
                    \ell_7=\ell_7^v+\ell_7^a \\[0.3cm]
                    \ell_{10}=\ell_{10}^v+\ell_{10}^a    
\end{array}\right.
&~~~~~~
\left\{\begin{array}{l}
    \ell_4^v=-\ell_5^v=1/[2\sqrt{2}\tilde{g}v\Lambda^{-1}]^2>0 \\[0.3cm]
    \ell_4^a=-\ell_5^a=\left[\eta^2(\eta^2-2)+16\tilde{\eta}^2\right]/
               [2\sqrt{2}\tilde{g}v\Lambda^{-1}]^2 \\[0.3cm]
    \ell_6^{v}=\ell_7^{v}=0 \\[0.3cm]
    \ell_6^{a}=-\ell_7^{a}
     =-\tilde{\eta}\left[4(3-\eta^2)\tilde{\eta}+(1-\eta^2)\eta\right]/
[2\sqrt{2}\tilde{g}v\Lambda^{-1}]^2 \\[0.3cm]
    \ell_{10}^{v}=\ell_{10}^{a}=0
   \end{array}\right.
\end{array}
\label{eq:VA-pattern}
}
\ee
where
\vspace*{-0.3cm}
\be 
{\small
~\eta = \dis\f{4\kappa_2}{4\kappa_2+\kappa_3}~,~~~~
\tilde{\eta} =\dis\f{2\kappa_2+4\tilde{\kappa}_2}
{\left(4\kappa_2+\kappa_3\right)+2\left(4\tilde{\kappa}_2+
\tilde{\kappa}_3\right)} - \f{2\kappa_2}{4\kappa_2+\kappa_3}~,
\label{eq:VA-para}
}
\ee
and $~\Lambda =\min(M_V,M_A)~$. 
At the leading order,    
$(M_V,M_A)\simeq \left(\tilde{g}v\sqrt{\kappa_1},~
\tilde{g}v\sqrt{\kappa_2+\kappa_3/4} \right)$,
after ignoring the SM gauge couplings $g$ and $g'$.
In (\ref{eq:VA-pattern}), the factor
{\small $~1/[\tilde{g}v\Lambda^{-1}]^2\simeq \kappa_1 (\Lambda /M_V)^2
=O(\kappa_1)$~} and all $SU(2)_c$-breaking terms depend on $\tilde{\eta}$~.
We see that the $SU(2)_c$-symmetric contribution 
from the axial-vector boson interactions to {\small ~$\ell_4^a=-\ell_5^a$~ }
becomes negative for {\small $~|\eta | <\sqrt{2}~$}, 
while the summed contribution {\small $~\ell_4=-\ell_5
=\left[ (\eta^2-1)^2+16\tilde{\eta}^2 \right]/
[2\sqrt{2}\tilde{g}v\Lambda^{-1}]^2\geq 0~$.}
The deviation of $\eta$ and/or $\tilde{\eta}$ 
from $\eta (\tilde{\eta}) =0$ represents the {\it non-QCD-like}
EWSB dynamics. For instance, in the case of 
$\tilde{g}=3$  and {\small $\{\eta,\tilde{\eta}\} =\{1.5,-0.25\}$,} we have 
{\small $~\ell_4=-\ell_5=2.35,~\ell_6=-\ell_7=-0.60,
 ~\ell_{10}=0$} for $\Lambda =2$~TeV.

\noindent
\underline{\bf Heavy Fermion Doublet}

Consider a simple
model with two heavy chiral fermions in the fundamental representation
of a new strong $SU(N)$ gauge group. They form a left-handed weak
doublet $(U_L,D_L)^T$ and right-handed singlets $(U_R, D_R)$. 
The small mass-splitting of fermions $U$ and $D$
breaks the $SU(2)_c$ and is characterized by the parameter
$~\omega =1-\left(M_D/M_U\right)^2~$. The anomaly-cancellation is ensured
by assigning the $U, D$ electric charges as
$(+\f{1}{2},-\f{1}{2})$~.
By taking $(U, D)$ as the source of the EWSB,
the $W,Z$ masses can be generated by heavy fermion loops. The new
contributions
to the quartic gauge couplings of $W/Z$ come from the {\it non-resonant}
$(U, D)$ box-diagrams\footnote{These box-diagrams were computed for the 
SM fermions in Ref.~\cite{box}.}.
The leading results in the $1/M_{U,D}$ and $\omega$ expansions
are derived as follows:
\be
{\small
\dis\ell_4^f=-2\ell_5^f=\left(\f{\Lambda}{4\pi v}\right)^2\f{N}{12}~;~~~
\dis\ell_6^f=-\ell_7^f=
-\left(\f{\Lambda}{4\pi v}\right)^2\f{7N}{240}\omega^2~,~~~
\dis\ell_{10}=0~;
\label{eq:fermion-pattern}
}
\ee
where $~\Lambda =\min(M_U,M_D)$.
For a model with $N=4$ and $(M_U,M_D)=(3.1,3.0)$~TeV,
{\it i.e.}, $\Lambda =3$~TeV and $\omega =0.063$, we have 
$~\ell_4^f=-2\ell_5^f\simeq 0.33,~$ and 
$~\ell_6^f=-\ell_7^f\simeq 0~$.

In summary, the typical values of the QGCs ($\ell_n$'s) are
expected to be around $O(1)$,
and different models of the strongly-interacting EWSB
sector result in different patterns of these parameters.
In the following sections we study the direct test of the QGCs
at the high energy linear colliders and show how the bounds can be improved 
in order to sensitively discriminate different models.

\vspace{0.5cm}
\noindent
{\normalsize\bf  3.~Testing Quartic Gauge Boson Couplings
via $WWZ/ZZZ$-Production}
\vspace{0.25cm}

While the CERN Large Hadron Collider (LHC) 
may provide the direct test on these new quartic 
gauge couplings (QGCs) through $WW$ fusion processes, the large
SM backgrounds make the experimental measurements difficult \cite{B-Y}.
The corresponding studies for the fusion processes
at the TeV $e^\pm e^-$ linear colliders
show that the complementary information
can be obtained \cite{han-lc,lc-vv}. 
On the other hand, the triple gauge boson production processes
\be
e^+e^- ~\longrightarrow~ W^+W^-Z,~~ ZZZ
\ee
have large cross sections just above the threshold
\cite{vvv-han} and may prove to be useful for probing the
QGCs \cite{vvv-other},  which contribute to the
signal diagrams involving $s$-channel $Z$-boson.
In this section,
we make a systematic analysis on the above processes and 
demonstrate the crucial role of the polarized $e^\mp$ beams. 

Because of the relatively clean experimental environment
at the LC, we identify the final state
$W/Z$'s via their hadronic dijet modes. Due to the
limited calorimeter energy resolution, the misidentification
probability of $W$ versus $Z$ should be considered~\cite{han-lc}.
To increase the statistics, we also add the clean channel
$Z \rightarrow \mu^-\mu^+$. To avoid other potential backgrounds of 
the type  $e^-e^+\to eeZZ,eeWW$, the electron-pair mode of $Z$-decay is  
not included.  We find that the detection
efficiencies for $WWZ$ and $ZZZ$ final states
are about 18.4\% and 16.8\%, respectively. 
It turns out that the $t$-channel $\nu_e$ exchange diagrams
in $e^-e^+\rightarrow WWZ$ production pose a large SM
background to the QGC signal, and they 
can hardly be suppressed by simple
kinematic cuts alone (cf. Fig.~1).  However, we note that  
such a type of
background involves the left-handed $W$-$e$-$\nu$ coupling and thus
can be very effectively suppressed by using the right(left)-handed
polarized $e^-(e^+)$ beam.
The highest sensitivity is reached by maximally polarizing
{\it both} $e^-$ and $e^+$ beams.
This is shown in Fig.~1 for distributions of the invariant mass
$M_{WW}$, transverse momentum $p_T(Z)$ and
$\cos\theta (e^-Z)$ at the 500~GeV LC,
without/with beam polarizations. To enhance the
signal-to-background ratio, we also impose some kinematical
cuts as indicated by the vertical lines in each panel of Fig.~1.

The crucial roles of the beam polarization and the higher
collider energy for the $WWZ$-production are
demonstrated in Fig.~2a, where $\pm 1\sigma$ exclusion
contours for $\ell_4$-$\ell_5$ are displayed at $\sqrt{s}=0.5,~ 0.8,~ 
1.0$ and $1.6$~TeV,  respectively.
The beam polarization has much less impact on the $ZZZ$ mode, due to  
the almost axial-vector type $e$-$Z$-$e$ coupling. Including the same
polarizations as in the case of the $WWZ$ mode, we find about 
$10-20\%$ improvements on the bounds from the $ZZZ$-production.
Assuming the two beam polarizations ($90\% ~e^-$ and $65\% ~e^+$),
we summarize the final $\pm 1\sigma$ bounds
for both $ZZZ$ and $WWZ$ channels and their combined $90\%$~C.L. contours
for $0.5$~TeV with $\int {\cal L}=50$~fb$^{-1}$ in Fig.~2b
(representing the {\it first direct probe} at the
LC) and for $1.6$~TeV with $\int {\cal L}=200$~fb$^{-1}$ in Fig.~2c 
(representing the {\it best} sensitivity gained
from the final stage of the LC with energy around $1.5/1.6$~TeV).
We see that, at the $90\%$~C.L. level, the bounds on $\ell_4$-$\ell_5$
at 0.5~TeV are within $O(10-20)$, while at 1.6~TeV they sensitively reach 
$O(1)$. The ellipses for the $WWZ$ final state in $\ell_4$-$\ell_5$ plane
are identical to those in $\ell_6$-$\ell_7$ plane, while the bands 
for the $ZZZ$ final state in $\ell_6$-$\ell_7$ plane 
become tighter due to a factor of 2 enhancement from the $4Z$-interaction
vertex.  $\ell_{10}$ only contributes to $ZZZ$ final state and can be
probed at the similar level.
The new physics cutoff is chosen as $~\Lambda =2$~TeV
in our plots and the numerical results for other values of
$~\Lambda~$ can be obtained by simple scaling.
In the above, the total rates are used to derive the numerical bounds.
We have further studied the possible improvements by including different
characteristic distributions (cf. Fig.~1), 
but no significant increase of the
sensitivity is found for the above processes.

Finally, a parallel analysis to Fig.~2b-c has been carried
out for the situation without $e^+$ beam polarization
(with $e^-$ polarization as before). For a two-parameter 
($\ell_{4,5}$) study, the $90\%$~C.L. results are compared as folllows:
\be
{\small 
\begin{array}{lcc}
{\rm at~0.5~TeV:}~~~ & 
-12~(-18)\leq \ell_4 \leq 21~(27), ~&~ -17~(-22)\leq \ell_5 \leq 9.5~(15);
\\[0.18cm]
{\rm at~1.6~TeV:}~~~ & 
-0.50~(-0.67)\leq \ell_4 \leq 1.5~(1.7), 
~&~ -1.3~(-1.5)\leq \ell_5 \leq 0.36~(0.58);
\end{array}
}
\label{eq:e+pol-compare}
\ee
where the numbers in the parentheses denote the bounds from polarizing the
$e^-$-beam alone.  The comparison in (\ref{eq:e+pol-compare}) 
shows that without $e^+$-beam polarization, 
the sensitivity will decrease by about $15\%-60\%$. 
Therefore, making use of the possible $e^+$-beam polarization with a degree 
around $65\%$ will certainly be beneficial.

\vspace{0.4cm}
\noindent
{\normalsize\bf 4.~Interplay of $WWZ/ZZZ$-Production and $WW$-Fusion }
\vspace{0.2cm}

The scattering amplitudes of the
longitudinally polarized $WW\to WW$ fusion have the
highest power dependence on the scattering energy $E$, 
while the $s$-channel $WWZ/ZZZ$ production at higher energies
suffers a reduction factor of $(v/E)^2$
relative to that of the fusion processes \cite{global}.
When the collider energy is reduced by half
(from $1.6$~TeV down to $800$~GeV),
the sensitivity of the $WW$-fusion decreases by about a factor of $20$
or more~\cite{lc-vv}. We therefore expect that $e^-e^+\to WWZ,ZZZ$
are more important at the earlier phase of the linear collider
and will be competitive with and complementary to the $WW$-fusion for 
later stages with energies $\sim$$0.8-1$~TeV. At even higher energies
$\sim$1.5/1.6~TeV, the fusion processes will take over due to their
higher energy dependence. 
The following analysis reveals 
that even at 1.6~TeV, $e^-e^+\to WWZ$ plays a crucial role for probing
the $SU(2)_c$-breaking parameters $(\ell_6,\ell_7)$.

For $WW$-fusion processes, there are five useful channels to consider:
\be
\begin{array}{lll}
 {\rm Full~process:} &  {\rm Subprocess:} &  {\rm
Parameter~dependence:}\\[0.26cm]
e^-e^+ \to \nu\bar{\nu}W^-W^+ ~, & (W^-W^+\to W^-W^+),~
& (\ell_{4,5})~, \\[0.13cm]
e^-e^- \to \nu\bar{\nu}W^-W^- ~, & (W^-W^-\to W^-W^-),
& (\ell_{4,5})~; \\[0.22cm]
e^-e^+ \to \nu\bar{\nu}ZZ ~,     & (W^-W^+\to ZZ),
& (\ell_{4,5};~\ell_{6,7})~, \\[0.13cm]
e^-e^+ \to e^\pm\nu W^\mp Z ~,~~& (W^\mp Z\to W^\mp Z ),
& (\ell_{4,5};~\ell_{6,7})~,  \\[0.13cm]
e^-e^+ \to e^-e^+ ZZ ~,~~& (ZZ\to ZZ),
& ([\ell_4+\ell_5]+2[\ell_6+\ell_7+\ell_{10}])~. 
\end{array}
\label{eq:fusion}
\ee
The first two processes in (\ref{eq:fusion}) only involve
$\ell_{4,5}$ and thus can provide a clean test on them.
Including the third and fourth reactions, one may further
probe $\ell_{6,7}$. Finally the fifth channel provides
information on $\ell_{10}$. However,
the realistic situation is more involved.
First of all, the $WZ$-channel has large $\gamma$-induced backgrounds
$e\gamma \rightarrow \nu WZ$ and $\gamma\gamma \rightarrow WW$. Secondly,
the small electron neutral-current coupling heavily suppresses the total 
rate of the $ZZ \rightarrow ZZ$ process.
Consequently, the sensitivity-bound on  $\ell_6$-$\ell_7$ from the
$e\nu WZ$ channel is significantly weaker than that from the $\nu\bar{\nu}ZZ$
channel \cite{lc-vv}, as illustrated in Fig.~3.
On the other hand, the triple gauge boson process $e^-e^+\to WWZ$ provides
complementary information. Fig.~3a demonstrates the
interplay of $WW$-fusion and $WWZ$-production for
discriminating the $SU(2)_c$-breaking parameters $\ell_{6}$-$\ell_7$ 
at a 1.6~TeV LC with 200 fb$^{-1}$ luminosity.
The sensitivity of $WWZ$ channel is shown to be comparable with
$\nu\bar\nu ZZ$ fusion channel in probing $\ell_{6,7}$. Here the bound
from $\nu eWZ$ channel is relatively too weak to be useful.
Including the  $WWZ$ channel,
the bound on $\ell_{6,7}$ is improved by about a factor of 2 and
thus reaches the same level as that of $\ell_{4,5}$ 
derived from the $\nu\bar{\nu} W^-W^+$ and
$\nu\nu W^-W^-$ channels \cite{lc-vv}.
To constrain $\ell_{10}$, both $ZZZ$ and $eeZZ$ channels are available.
Assuming that $\ell_{4,5;6,7}$ are constrained by the processes mentioned
above, we set their values to be zero (the reference point) for simplicity 
and define the statistic significance
$~S=|{\cal N}-{\cal N}_{0}|/\sqrt{{\cal N}_{0}}~$ which is
a function of $\ell_{10}$. (Here ${\cal N}$ is the
total event-number while ${\cal N}_0$ is the number at $\ell_{10}=0$.)
As shown in Fig.~3b, at 1.6~TeV, the sensitivity of $e^-e^+\to eeZZ$ 
for probing $\ell_{10}$ is better than that of $e^-e^+\to ZZZ$.

\vspace*{0.5cm}
\noindent
{\normalsize\bf 5.~Conclusions}\\[0.3cm]
After analyzing the different patterns of the quartic
gauge boson couplings
in connection with the representative strongly-interacting models,
we study the sensitivities in probing both the $SU(2)_c$-symmetric and  
-breaking QGCs via $WWZ/ZZZ$-production at the LCs.
We summarize in Table~1 the combined $90\%$~C.L. sensitivity bounds on  
the QGCs from $WWZ$ and $ZZZ$ channels
for typical energies and luminosities of the LCs, which  
are to be compared with the estimated {\it indirect} LEP-bounds in (2).
We further analyze the interplay of the $WWZ/ZZZ$-production with the
$WW$-fusion mechanism for achieving an improved determination of all the 
five QGCs.  The important roles of both the polarized $e^-$ and
$e^+$ beams for the $WWZ$-production are revealed and analyzed.

The first direct probe on these QGCs will come from the early 
phase of the LC at 500~GeV, where
the $WW$-fusion processes are not useful.
The two mechanisms become more competitive and
complementary at energies $\sqrt{s}\sim 0.8-1$~TeV. 
At a later stage of the LC, $\sqrt{s}=1.6$~TeV, 
the $90\%$~C.L. one-parameter bounds from the fusion 
processes become very sensitive, for $\Lambda =2$~TeV:
\be
{\small 
\begin{array}{c}
-0.13 \leq \ell_4 \leq 0.10~,~~~~-0.08 \leq \ell_5 \leq 0.06 ~;\\
-0.22 \leq \ell_6 \leq 0.22~, ~~~~-0.12 \leq \ell_7 \leq 0.10 ~,~~~~
-0.21 \leq \ell_{10} \leq 0.21 ~;
\end{array}
\label{eq:fusion-bound}
}
\ee
obtained for $\int{\cal L}=200$ fb$^{-1}$ 
with a $90\%$ ($65\%$) polarized $e^-$($e^+$) beam.
The bounds on $\ell_{4,5}$ are about a factor of $3\sim 6$ stronger 
than that from $WWZ/ZZZ$-modes (cf. Table~1);
while the bounds on $\ell_{6,7,10}$ are comparable.
For a complete multi-parameter analysis, the $WWZ$-channel is crucial for
determining $\ell_6$-$\ell_7$ even at a 1.6~TeV LC. \\[0.2cm]
{\small Table~1: Combined $90\%$~C.L. bounds on $\ell_{4-10}$
from $WWZ/ZZZ$-production. For simplicity, we\\[-0.26cm]
set one parameter to be nonzero at a time. The bound on $\ell_{10}$ 
comes from $ZZZ$-channel alone.\\[-0.45cm]
} 
\begin{center}
\tabcolsep 1pt
{\small
\begin{tabular}{c||c|c|c|c}
\hline\hline
&&&&\\[-0.2cm]
$\sqrt{s}$~(TeV) & 0.5 & 0.8 & 1.0 & 1.6 \\[0.15cm]
\hline
&&&&\\ [-0.36cm]
$\int{\cal L}$~(fb$^{-1}$) & 50
& 100 & 100 & 200  \\ [0.2cm]
\hline\hline
&&&&\\[-0.33cm]
& $~-9.5\leq\ell_4\leq 11.7~$   & $~-2.7\leq\ell_4\leq 3.2~$
& $~-1.7\leq\ell_4\leq 2.0~$   & $~-0.50\leq\ell_4\leq 0.58~$ \\[0.2cm]
$WWZ/ZZZ$~
& $-9.8\leq\ell_5\leq ~8.9$   & $-3.1\leq\ell_5\leq 2.3$
& $-1.9\leq\ell_5\leq 1.4$   & $-0.54\leq\ell_5\leq 0.36$ \\[0.2cm]
Bounds
& $-5.0\leq\ell_6\leq 5.8$          & $-1.5\leq\ell_6\leq 1.6$
& $-0.95\leq\ell_6\leq 1.0$          & $-0.28\leq\ell_6\leq 0.28$  
\\[0.2cm]
(at 90$\%$C.L.)~
& $-5.0\leq\ell_7\leq 5.7$          & $-1.5\leq\ell_7\leq 1.5$
& $-0.95\leq\ell_7\leq 0.92$          & $-0.28\leq\ell_7\leq 0.26$\\[0.2cm]
& $-4.3\leq\ell_{10}\leq 5.2$       & $-1.4\leq\ell_{10}\leq 1.4$
& $-0.83\leq\ell_{10}\leq 0.88$        & $-0.26\leq\ell_{10}\leq  
0.26$ \\[0.2cm]
\hline
&&&&\\ [-0.36cm]
Range of $|\ell_n|$
& $\leq O(4\sim 10)$    &   $\leq O(1\sim 3)$
& $\leq O(0.8\sim 2)$  &   $\leq O(0.3\sim 0.6)$\\[0.3cm] 
\hline\hline
\end{tabular}
}
\end{center}

\vspace{0.8cm}
\noindent
{\normalsize\bf Acknowledgments}~~~ 
We thank E.~Boos, R.~Casalbuoni, D.~Dominici, 
G.~Jikia, Y.P.~Kuang, H.Y.~Zhou, I.~Kuss,
A.~Likhoded, C.R.~Schmidt, G.~Valencia, and P.M.~Zerwas for  
valuable discussions.
TH is supported by U.S.~DOE under contract DE-FG03-91ER40674
and the Davis Institute for High Energy Physics;
HJH by the AvH of Germany; CPY and HJH by the U.S.~NSF under  
grant PHY-9507683.

\vspace{0.3cm}
\noindent 
{\bf Note added}: 
After the submission of this paper, we received a paper from O.J.P. Eboli
{\it et al.} \cite{added-eboli} who also studied the $WWZ/ZZZ$ production
for probing the quartic gauge boson couplings. With different cuts and beam
polarization choice, they obtained weaker bounds than ours. They reached a
similar conclusion on the role of the $e^-$-polarization for the $WWZ$-channel.

\pagebreak

\addtolength{\topmargin}{0.6cm}

%
%
\begin{figure}[p]
\begin{center}
\epsfig{file=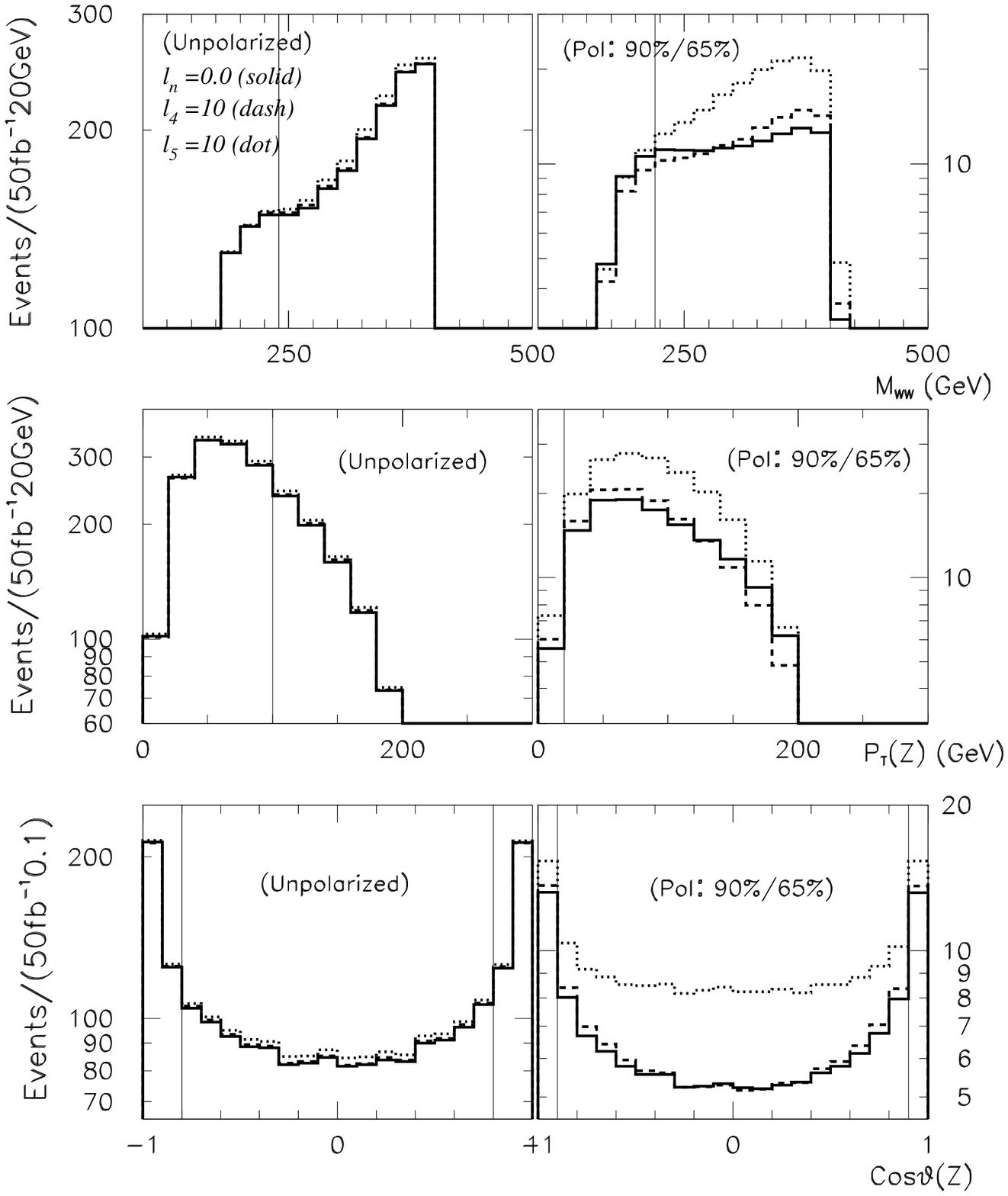,height=20cm,width=16cm}
\end{center}
\vspace{-1.5cm}
\caption{Kinematical distributions and cuts 
(indicated by the vertical lines in each panel) 
for $e^-e^+\rightarrow WWZ$ at
0.5~TeV with an integrated luminosity 50~fb$^{-1}$.
Results with $\ell_n=0$ (solid), $\ell_4=10$ (dashes)
and $\ell_5=10$ (dotted) are shown. The left and right
panels compare the results without and with beam polarizations.
}
\label{fig:figwwz500}
\end{figure}

%
%
\addtolength{\topmargin}{-1.5cm}
\addtolength{\textheight}{1.5cm}
\null\vspace*{-1.5cm}
\begin{figure}[T]
\hspace*{-2cm}
\epsfig{file=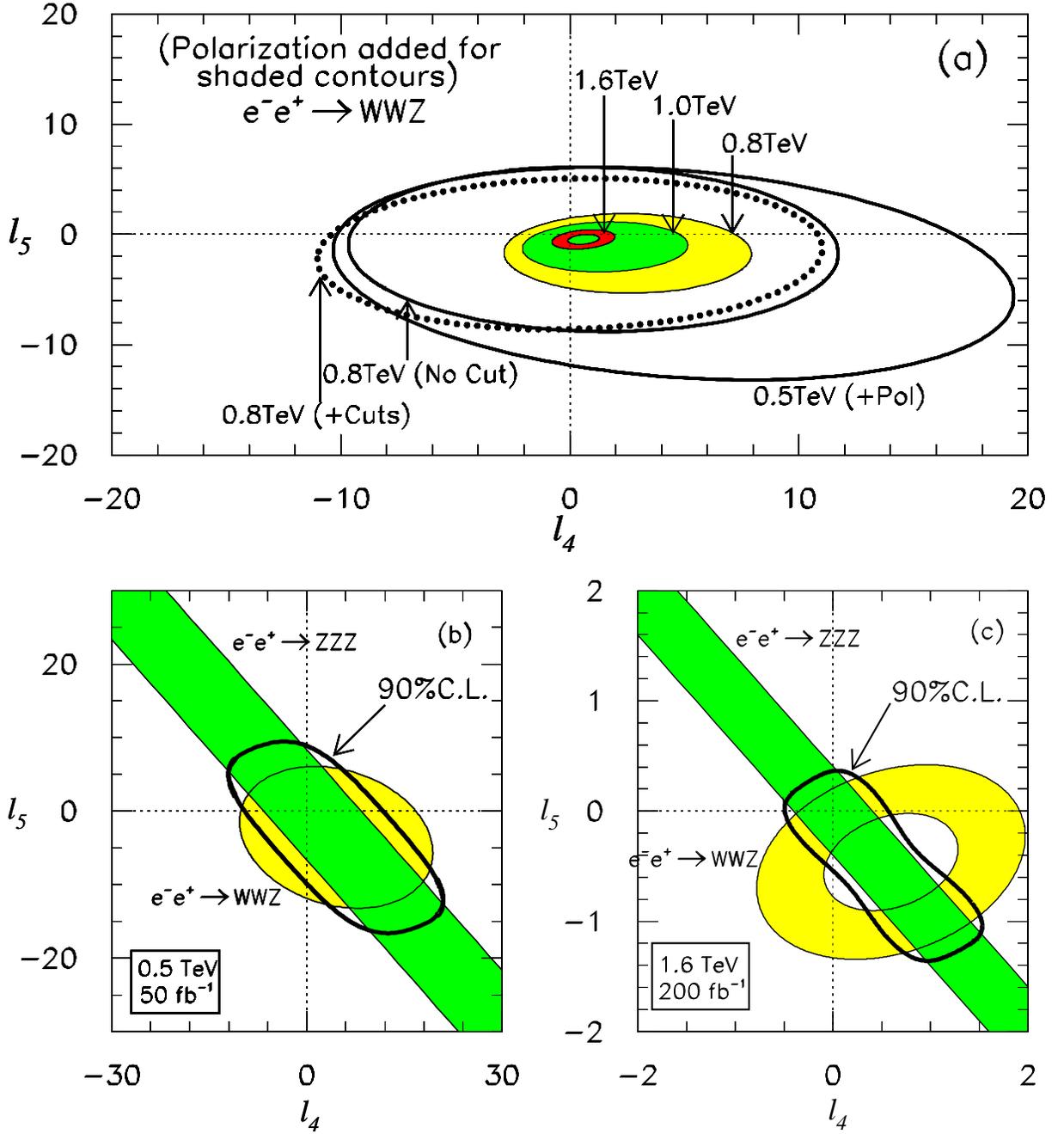,height=21cm} 
\vspace*{-2.1cm}
\caption{Probing $\ell_4$-$\ell_5$ via $WWZ$ and $ZZZ$ production processes.
The roles of the polarization and the higher collider energy for
$e^-e^+\to WWZ$ are shown by the $\pm\,1\sigma$ exclusion contours  
in (a).
The integrated luminosities used here are 50~fb$^{-1}$ (at 500~GeV),
100~fb$^{-1}$ (at 800~GeV) and 200~fb$^{-1}$ (at 1.0 and 1.6~TeV).
In (b) and (c), the $\pm\,1\sigma$ contours are displayed for $ZZZ/WWZ$
final states at $\sqrt{s}=$0.5 and 1.6~TeV respectively,
with two beam polarizations ($90\% ~e^-$ and $65\% ~e^+$);
the thick solid lines present the combined bounds at $90\%$~C.L.
}
\label{figvvv-bound}
\end{figure}

%
\null\vspace*{-1.5cm}
\begin{figure}[h] 
\hspace*{-1.4cm}
\epsfig{file=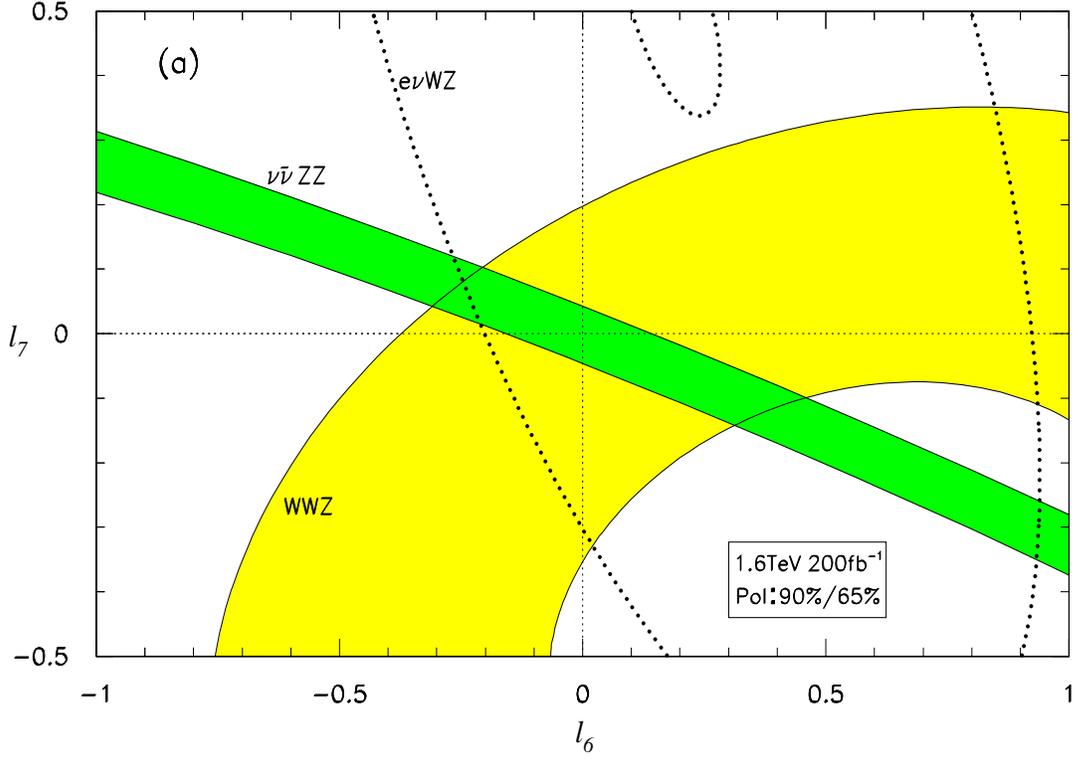,height=12.5cm,width=17cm}\\[-1.5cm]
\hspace*{-1.2cm}
\epsfig{file=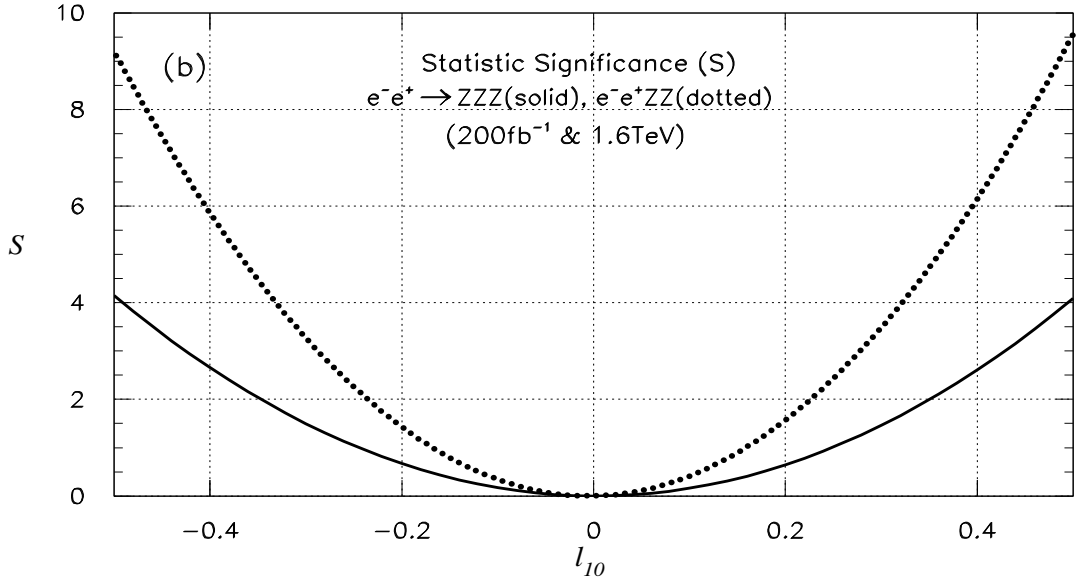,height=10.0cm,width=17cm}
\vspace{-1.2cm}
\caption{Interplay of the $WW$-fusion and $WWZ/ZZZ$-production for
discriminating 
$\ell_{6}$-$\ell_7$ and $\ell_{10}$
at $~\sqrt{s}=$1.6~TeV with $\int{\cal L}=$200~fb$^{-1}$:
(a). $\pm\,1\sigma$ exclusion contours for  $~e^-e^+ \to  
\nu\bar{\nu}ZZ,~
e^+\nu W^-Z/e^-\bar{\nu}W^+Z$,~ and $e^-e^+\to WWZ$~
with polarizations ($90\% ~e^-$ and $65\% ~e^+$).
(b). Statistic significance versus $\ell_{10}$
for $e^-e^+\to ZZZ,~e^-e^+ZZ$ (with unpolarized $e^\mp$ beams). }
\label{fig-interplay}
\end{figure}
\clearpage

\end{document}
\end